\documentclass[a4paper,11pt]{article}
\usepackage{pos}

\title{Flavour Physics with SuperIso}
\ShortTitle{Flavour Physics with SuperIso}

\author*[a]{S. Neshatpour}
\author[a,b]{F. Mahmoudi}

\affiliation[a]{Univ Lyon, Univ Claude Bernard Lyon 1, CNRS/IN2P3, IP2I Lyon, UMR 5822, F-69622, Villeurbanne, France}

\affiliation[b]{Theoretical Physics Department, CERN, CH-1211 Geneva 23, Switzerland\\~}

\emailAdd{siavash.neshatpour@univ-lyon1.fr }
\emailAdd{nazila@cern.ch}

\abstract{%
We describe \texttt{SuperIso v4.1},  a public code for calculation of flavour observables, in the Standard Model and for New Physics scenarios in a model-independent way as well as for specific beyond the SM models such as two Higgs doublet models (2HDM), minimal supersymmetric Standard Model (MSSM) and next-to-minimal supersymmetric Standard Model (NMSSM).
}

\FullConference{%
  Tools for High Energy Physics and Cosmology - TOOLS2020\\
  2-6 November, 2020\\
  Institut de Physique des 2 Infinis (IP2I), Lyon, France}

\begin{document}

\maketitle

\section{Introduction}
The Standard Model of Particle Physics has had a remarkable agreement with experimental measurements. 
Nonetheless, there are strong reasons to believe that there is New Physics (NP) beyond the Standard Model.
Direct searches for new particles predicted from these models are ongoing at particle accelerators.
Complementary to these direct searches are indirect searches for new particles which can contribute as virtual states 
both in Flavour Changing Charged Current (FCCC) transitions as well as in Flavour Changing Neutral Current (FCNC) processes.
Flavour observables are effective indirect probes of NP, especially in FCNC processes as they
only occur through virtual loops in the SM and hence are suppressed. Comparison of the theoretical
predictions for observables relevant to these decays with their experimental measurements
could unveil the presence of heavy particles propagating in the loops - particles not accounted for in the Standard Model.
This approach was used to estimate the large top quark mass before it was actually measured at the Tevatron, and ever since has become
one of the main approaches for searching for the effect of New Physics at low energies.

\texttt{SuperIso} is a public C program~\cite{Mahmoudi:2007vz,Mahmoudi:2008tp,Mahmoudi:2009zz} dedicated to calculating flavour physics observables 
in the SM and in beyond the SM scenarios. Besides flavour observables which are described in the next section other observables such as the muon anomalous moment $a_\mu$,
electroweak precision tests; oblique parameters ($S, T, U$), $\rho, \Gamma_z$ are implemented in \texttt{SuperIso}. 
Moreover, direct search limits from LEP and Tevatron are also available. 
Furthermore in a separate code, \texttt{SuperIso Relic}~\cite{Arbey:2009gu,Arbey:2011zz,Arbey:2018msw}, which is an extension of \texttt{SuperIso}, observables related to dark matter direct and indirect detections can be considered.

\section{Flavour observables}
\label{sec:obs}
Flavour observables implemented in \texttt{SuperIso} can be categorized into two groups of
FCNC and  FCCC transitions. 

The FCNC processes include the branching ratio of radiative decays such as
the inclusive $B \to X_{s,d} \gamma$,  and  the exclusive $B \to K^* \gamma$ which are mostly sensitive to $C_7$, the Wilson coefficient of the radiative penguin operator. 
The branching ratios of the leptonic decays $K_{L,S} \to \ell^+ \ell^-$ and $B_{s,d} \to \ell^+ \ell^-$ are sensitive to the axial-vector Wilson coefficient $C_{10}$ as well as scalar and pseudo scalar operators putting strong constraints on models involving an extended Higgs sector.
For semileptonic FCNC decays besides the branching ratios, there are several angular observables available giving access to different combinations of the aforementioned Wilson coefficients as well as to the vector Wilson coefficient $C_9$. The semileptonic decays and their corresponding observables implemented in \texttt{SuperIso} are
\begin{itemize}
\item $B \to X_s \ell^+ \ell^-$: BR, $A_{\rm FB}$, zero-crossing $q_0^2 (A_{\rm FB})$ 
\item $B^{(+)} \to K^{*(+)} \ell^+ \ell^-$: BR, $A_{\rm FB}$, zero-crossing $q_0^2 (A_{\rm FB})$, $F_L, A_T^{i}, P_i^{(\prime)}, S_i, R_{K^*}$ 
\item $B_s \to \phi \ell^+ \ell^-$: BR, $P_i^{(\prime)}, S_i, R_{\phi}$ 
\item $B^{(+)} \to K^{(+)} \ell^+ \ell^-$: BR, $A_{\rm FB}, F_H, R_{K}$ 
\item $\Lambda_b^+ \to \Lambda \ell^+ \ell^-$: BR, $A_{\rm FB}, F_L$ 
\item $K^+ \to \pi^+ \nu \bar\nu$: BR
\item $K_L \to \pi^0 \nu \bar\nu$: BR 
\end{itemize}
where BR refers to the branching ratio, $A_{\rm FB}$ to the Forward-Backward asymmetry and the definition of the angular observables  $S_i$, and $P_i^{(\prime)}$ can be found in Refs.~\cite{Altmannshofer:2008dz,Descotes-Genon:2013vna}.  
$R_H$ is the lepton flavour universality observable BR$(B_{(s)} \to H \mu^+ \mu^-)$/BR$(B_{(s)} \to H e^+ e^-)$  with $H = K, K^*, \phi$  as proposed in Ref.~\cite{Hiller:2003js}.
Furthermore, the  $\Delta S = 2$ Wilson coefficients can be probed via the $B_{d,s}$ meson-mixing observables.  

In the FCCC category, the branching ratios of $B \to \ell \nu$, $B~\to~D^{(*)} \ell \nu$, $D_s \to \ell \nu, D \to \mu \nu$ 
and $K \to \mu \nu$ are among the most important observables, that are implemented in SuperIso.

\section{Installation}
The \texttt{SuperIso} code can be downloaded from \href{http://superiso.in2p3.fr}{http://superiso.in2p3.fr} which upon unpacking creates the main directory \texttt{superiso\_vX.X}\,.

The main directory includes thirteen main programs predicting observables in different models; \texttt{sm.c} calculates a few sample flavour observables in the SM, while 
\texttt{amsb.c, cmssm.c, cnmssm.c, gmsb.c, hcamsb.c, mmamsb.c, ngmsb.c, nnuhm.c, nuhm.c, thdm.c} make predictions in specific NP models (which require to be linked to one of the external spectrum generator programs \texttt{ISAJET~\cite{Paige:2003mg},\; NMSSMTools~\cite{Ellwanger:2005dv},\; SOFTSUSY~\cite{Allanach:2001kg},\; SPheno~\cite{Porod:2003um},\;  SuSpect~\cite{Djouadi:2002ze},\; 2HDMC~\cite{Eriksson:2009ws}}, etc.) and \texttt{flha.c} and \texttt{slha.c} predict observable values taking as input a file of FLHA and SLHA format, respectively (a sample \texttt{example.lha} input file in the SLHA format is included).   

In the main directory there are three programs calculating the  $\chi^2$  for a sample predefined set of flavour observables.  \texttt{sm\_chi2.c} calculates the $\chi^2$ for the SM, while \texttt{modelindep\_chi2.c} and \texttt{slha\_chi2.c} give the $\chi^2$ within a model-independent scenario taking sample Wilson coefficients as inputs and using a SLHA type input, respectively.

There is also a \texttt{README} file as well a \texttt{Makefile} file in the main directory where in the latter the compiler can be specified.  By default, the compiler is set to gcc in \texttt{Makefile}. If gcc is available on your system the code can be compiled by simply typing \texttt{make} in the terminal which creates \texttt{libisospin.a} in \texttt{src/}. The provided main programs can then be compiled by typing \texttt{make name} where
\texttt{name} is any of the sixteen mentioned main programs,  creating an executable program with the~\texttt{.x} extension (e.g. \texttt{make sm.c} creates \texttt{sm.x}).
Further information is available in the \texttt{README} file or the manual which can be downloaded from the \texttt{SuperIso} website \href{http://superiso.in2p3.fr/superiso4.1.pdf}{http://superiso.in2p3.fr/superiso4.1.pdf}.

\section{Global fits}
\label{sec:models}
\begin{figure}[!t]
\begin{center}
\includegraphics[height=0.35\textwidth]{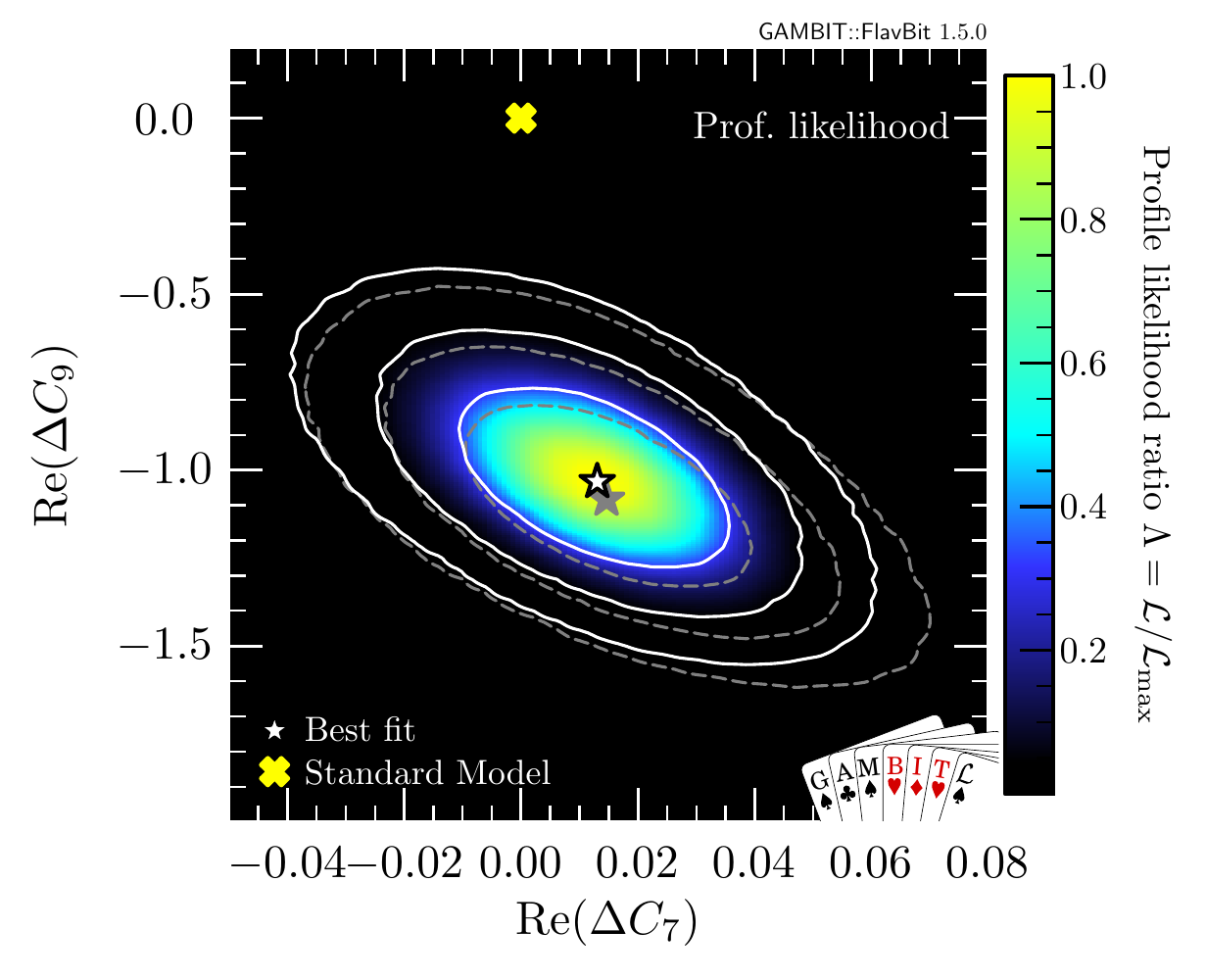}\qquad
\includegraphics[height=0.35\textwidth]{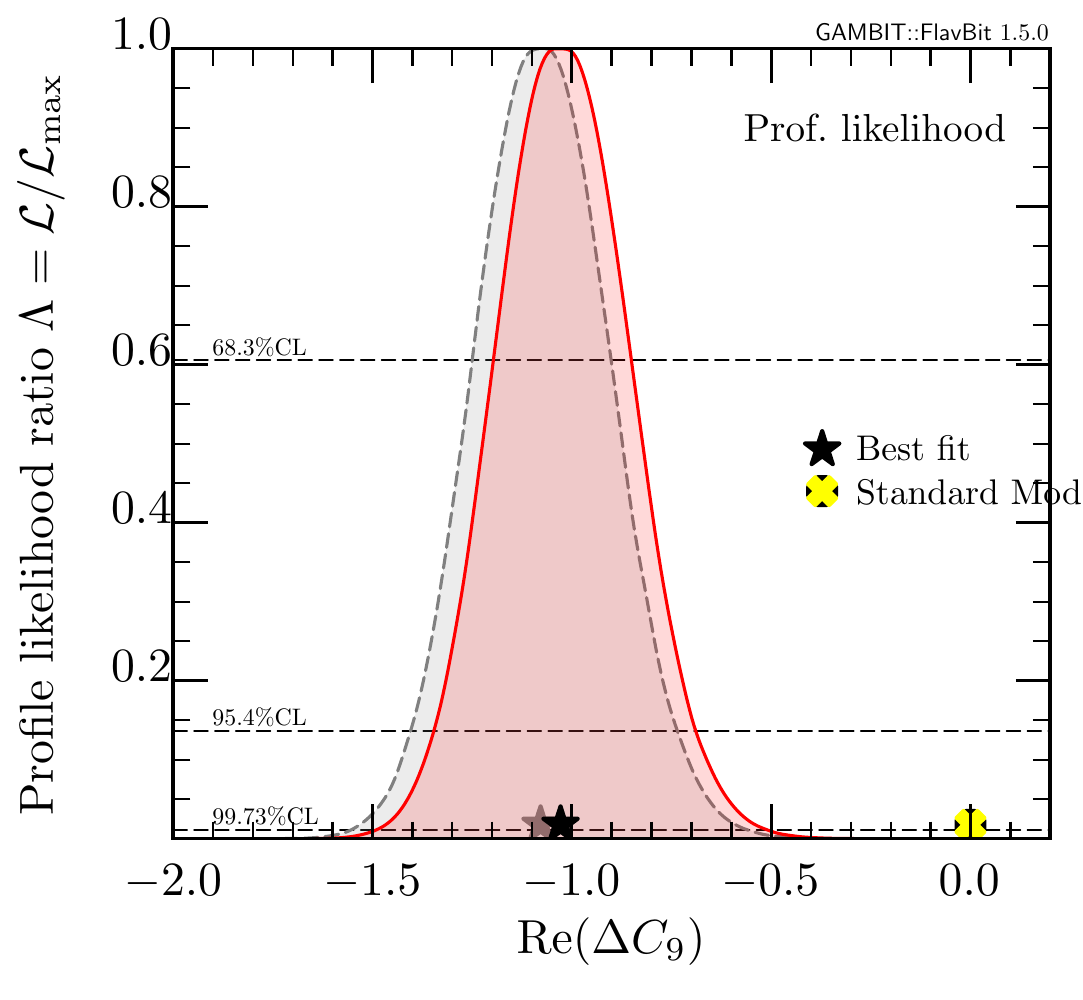}
\caption{Two- and one-dimensional profile likelihoods for ($C_7,C_{10}$) and $C_9$  from Ref.~\cite{Bhom:2020lmk}. 
The white (red) contour (curve) in the two (one)-dimensional result corresponds to the fit when the covariance matrix is recalculated for each point
while the the grey contour (curve) is given for when using the SM covariance matrix for all points.
\label{fig:CovDep}}
\end{center}
\end{figure}

In recent years, the strongest signs of New Physics have been observed in flavour physics, especially 
in flavour changing neutral current processes with quark level exchange $b \to s$ . 
However, the numerous flavour observables are interconnected via Wilson coefficients, and in order to  have a consistent
view of the implications of experimental measurements and possible anomalies in flavour physics observables, the global behaviour 
should be considered. This can be done by considering the $\chi^2$ of the chosen set of Wilson coefficients
\begin{align}
 \chi^2 =  \sum_{i,j =1}^{N} \left( O_i^{\rm th} - O_i^{\rm exp} \right)
 \; C_{ij}^{-1} \; \left( O_j^{\rm th} - O_j^{\rm exp} \right)\,,
\end{align}
where $O_{i}^{\rm th}$ and $O_{i}^{\rm exp}$ correspond to the theoretical prediction and experimental measurement of the $i$-th observable, respectively.   
$C_{i,j} $ denotes the $(i,j)$ element of the total covariance matrix which is the sum of the theoretical and experimental covariance matrices
\begin{align}
 C_{i,j}  = {\rm Cov}_{\rm th}(O_i, O_j) + {\rm Cov}_{\rm exp}(O_i, O_j)\,.
\end{align}
In \texttt{SuperIso}, for  ${\rm Cov}_{\rm exp}$ we take the covariance matrix provided by the experiment (when available) 
while for ${\rm Cov}_{\rm th}$ we calculate it assuming the nuisance parameters have Gaussian distribution 
and that the observables are linearly dependent on them~\cite{Arbey:2016kqi}, i.e.
the total variation of $O_i$ at first order is given~by
\begin{align}
 O_i = O_i^{(0)} \left( 1 + \sum_{a=1} \delta_a \,\Delta_{O_i}^{a} \right)\,,
\end{align}
where $O_i^{(0)}$ is the central value, $\delta_a$ denotes the variation of nuisance parameters $a$ and $\Delta_{O_i}^{a}$ is the relative variance of observable $O_i$ generated by the nuisance parameter. The correlation coefficient between the $O_i$ and $O_j$ observables is given by
\begin{align}
 \left( \Delta_{O_i, O_j} \right)^2 = \sum_{a,b} \rho_{ab} \, \Delta_{O_i}^a \, \Delta_{O_j}^b\,,
\end{align}
where $\rho_{ab}$ is the covariance between the nuisance parameters $a$ and $b$. The theoretical covariance matrix is then defined as 
\begin{align}
 {\rm Cov}_{\rm th} = 
\left( \begin{array}{cc}
 \left( \Delta_{O_i, O_i} \right)^2 \left( O_i^{(0)} \right)^2      &  \left( \Delta_{O_i, O_j} \right)^2 O_i^{(0)} O_j^{(0)}   \\
 \left( \Delta_{O_i, O_j} \right)^2  O_i^{(0)} O_j^{(0)}            &  \left( \Delta_{O_j, O_j} \right)^2 \left( O_j^{(0)} \right)^2 \end{array} \right).
\end{align}
By default in \texttt{SuperIso}, each nuisance parameter is varied within one standard deviation and the covariance matrix $C_{ij}$ is 
calculated for the SM and used for other models. While it is possible to calculate the covariance matrix for every New Physics point, when doing a 
large scan over the parameter space, it will be time-consuming as for each point it takes a few seconds when considering  around hundred observables.
A comparison between a 3-dimensional global fit to Wilson coefficients $\{C_7, C_9 , C_{10}\}$ when recalculating the covariance matrix for each point versus considering the one calculated for SM at all points has been done in Ref.~\cite{Bhom:2020lmk}  (see Fig.~\ref{fig:CovDep}) using \texttt{SuperIso} as part of \texttt{FlavBit/GAMBIT}~\cite{Workgroup:2017myk} indicating that using the covariance matrix calculated for the SM at all points is a reasonable approximation.

In \texttt{SuperIso} it is also possible to consider the nuisance parameters as free parameters that 
can be fitted. This is relevant for the case where there is a lack of precise knowledge on these parameters. Such fits have been done in Refs.~\cite{Chobanova:2017ghn,Arbey:2018ics,Hurth:2020rzx} for the power corrections that  contribute to $B \to K^* \mu^+ \mu^-$ observables, as the size of the hadronic corrections are not well-known and they can impact the significance of short-distance NP fits.

\begin{figure}[!t]
\begin{center}
\includegraphics[width=0.45\textwidth]{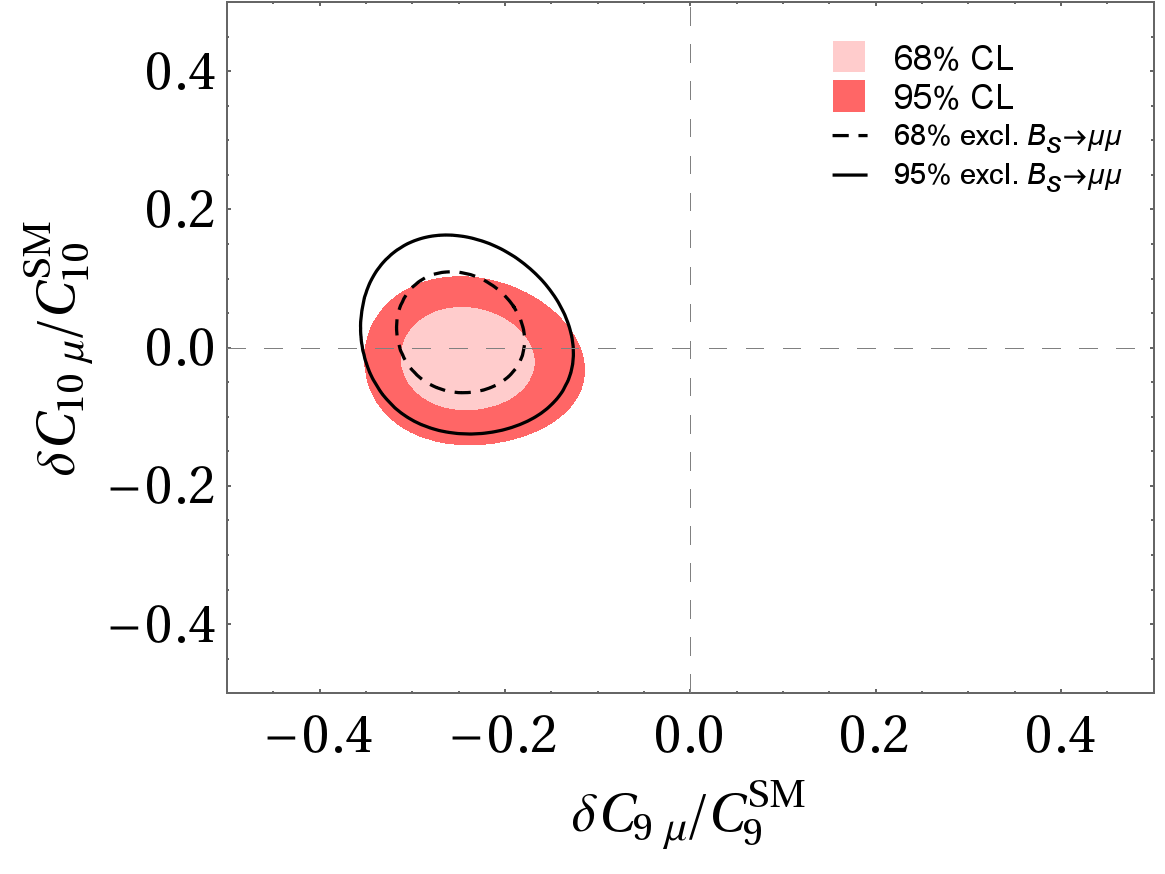}\qquad
\includegraphics[width=0.45\textwidth]{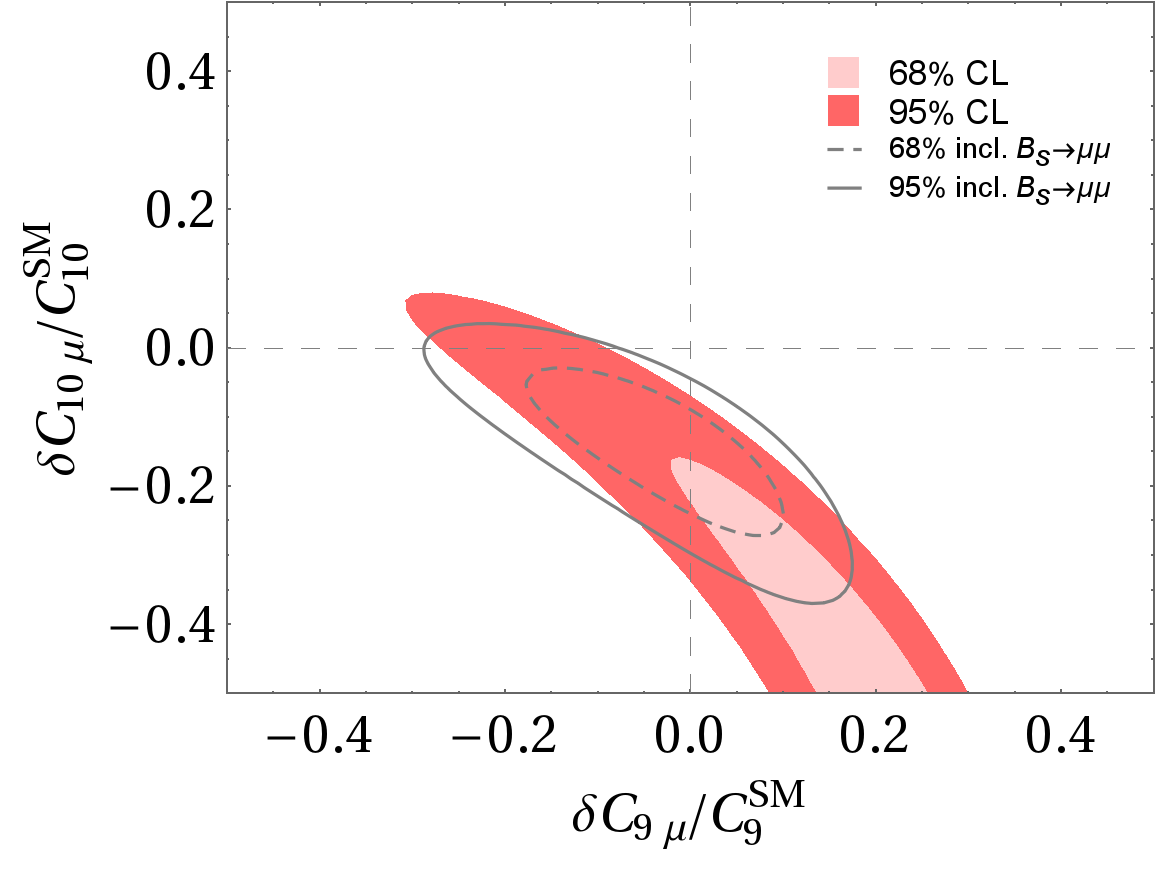}
\caption{Fit of $(C_{9}^\mu, C_{10}^\mu)$ to all observables except $R_K$ and $R_{K^*}$ on the left and to only the data on $R_K, R_{K^*}$ on the right
from Ref.~\cite{Arbey:2019duh}.
The red (light-red) contours correspond to 68\% (95\%) confidence level.  
The dashed gray (solid black) contours correspond to including (excluding) the data on $B_{s,d}\to \mu^+ \mu^-$.
\label{fig:Ratios_Rest}}
\end{center}
\end{figure}

Using a subset of observables the coherence of the implications of various measurements can be analysed.
In Fig.~\ref{fig:Ratios_Rest} Wilson coefficients fits for $(C_{9}^\mu, C_{10}^\mu)$ are given when considering two sets of observables; lepton flavour violating ratios $R_{K^{(*)}}$ and the rest of $b \to s \ell^+ \ell^-$ observables, indicating agreement at $2\sigma$ level~\cite{Arbey:2019duh}.
Another comparison of different sets of observables using \texttt{SuperIso} can be found in Ref.~\cite{Hurth:2020ehu} where the coherence 
of the angular observables of the neutral $B^{0} \to K^{*0} \mu^+ \mu^-$ decay~\cite{Aaij:2020nrf} and its charged counterpart, the $B^{+} \to K^{*+} \mu^+ \mu^-$ decay~\cite{Aaij:2020ruw} are studied.

\begin{figure}[!t]
\begin{center}
\includegraphics[width=0.32\textwidth]{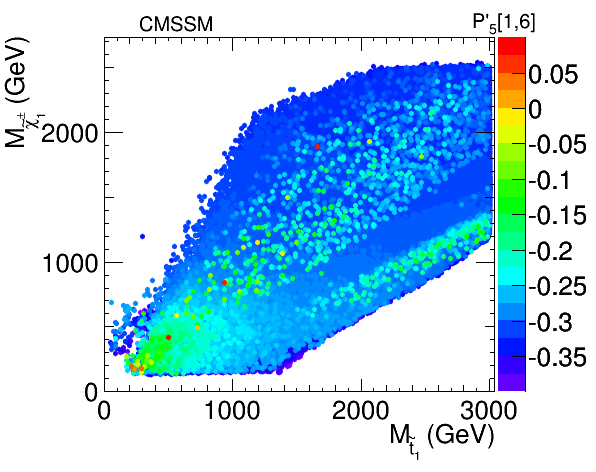}
\includegraphics[width=0.32\textwidth]{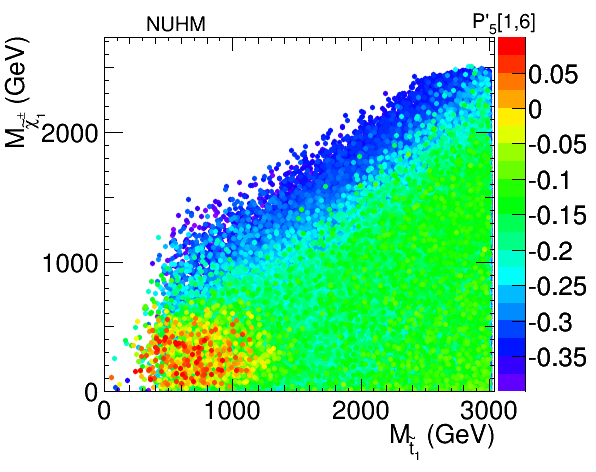}
\includegraphics[width=0.32\textwidth]{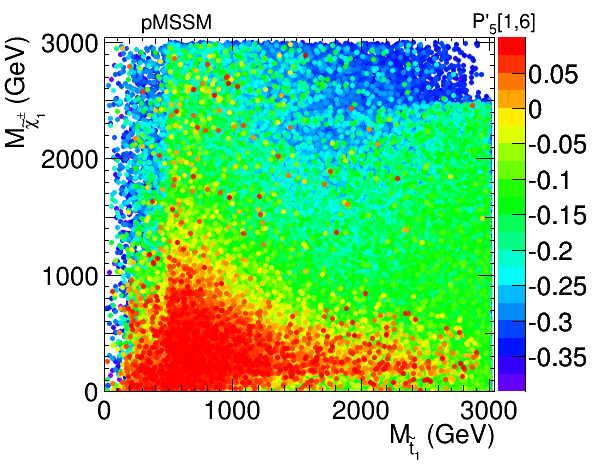}
\caption{$P_5^\prime (B \to K^* \mu^+ \mu^-)$  in the $[1,6]$ GeV$^2$ bin in the $( M_{\tilde{t}_1}, M_{\tilde{\chi}_1^\pm})$ plane for 
CMSSM on the left, NUMH in the middle and pMSSM on the right from Ref.~\cite{Mahmoudi:2014mja}. 
The most recent experimental measurement by LHCb~\cite{Aaij:2020nrf} for $P_5^\prime$ in the $[1.1,6]$ GeV$^2$ bin is $-0.144 \pm 0.068 \pm 0.026$
and its SM prediction is $-0.39 \pm 0.1$.
\label{fig:P5pMSSM}}
\end{center}
\end{figure}

It is also possible to consider global fits in  specific New Physics models or to check the effect of specific model parameters on individual observables.
In Fig.~\ref{fig:P5pMSSM} from Ref.~\cite{Mahmoudi:2014mja} the angular observable $P_5^\prime (B \to K^* \mu^+ \mu^-)$ (which is one of the main observables having tension with the SM predictions)
is given as a function of MSSM parameters the chargino and stop masses in the constrained MSSM models CMSSM, the Non
Universal Higgs Mass (NUHM) and the more general phenomenological MSSM (pMSSM).

Besides the specific models for which main programs are available in \texttt{SuperIso} where the Wilson coefficients are
automatically calculated once the model parameters are provided through spectrum generators, \texttt{SuperIso} 
can calculate flavour observables for any other New Physics models if the relevant Wilson coefficients are given. 
In the near future there is going to be a direct interface to \texttt{MARTY}~\cite{Uhlrich:2020ltd}  which is a
C++ public code capable of calculating Wilson coefficients for generic beyond the SM scenarios.

\section{Conclusions}
Flavor observables are powerful probes of New Physics models and currently offer the
most promising signs of New Physics.  Numerous flavour observables are implemented in \texttt{SuperIso} which can be used to 
study beyond the Standard Model scenarios both in a model-independent way as well as in specific 
models such as 2HDM and Superysymmetric models.

\acknowledgments
The speaker, SN, is grateful to the TOOLS 2020 organisers for their invitation to present this~work. 

\bibliographystyle{JHEP}
\bibliography{Tools2020}

\providecommand{\href}[2]{#2}\begingroup\raggedright\begin{thebibliography}{10}

\bibitem{Mahmoudi:2007vz}
F.~Mahmoudi, \emph{{SuperIso: A Program for calculating the isospin asymmetry
  of $B \to K^* \gamma$ in the MSSM}},
  \href{https://doi.org/10.1016/j.cpc.2007.12.006}{\emph{Comput. Phys. Commun.}
  {\bfseries 178} (2008) 745}
  [\href{https://arxiv.org/abs/0710.2067}{{\ttfamily 0710.2067}}].

\bibitem{Mahmoudi:2008tp}
F.~Mahmoudi, \emph{{SuperIso v2.3: A Program for calculating flavor physics
  observables in Supersymmetry}},
  \href{https://doi.org/10.1016/j.cpc.2009.02.017}{\emph{Comput. Phys. Commun.}
  {\bfseries 180} (2009) 1579}
  [\href{https://arxiv.org/abs/0808.3144}{{\ttfamily 0808.3144}}].

\bibitem{Mahmoudi:2009zz}
F.~Mahmoudi, \emph{{SuperIso v3.0, flavor physics observables calculations:
  Extension to NMSSM}},
  \href{https://doi.org/10.1016/j.cpc.2009.05.001}{\emph{Comput. Phys. Commun.}
  {\bfseries 180} (2009) 1718}.

\bibitem{Arbey:2009gu}
A.~Arbey and F.~Mahmoudi, \emph{{SuperIso Relic: A Program for calculating
  relic density and flavor physics observables in Supersymmetry}},
  \href{https://doi.org/10.1016/j.cpc.2010.03.010}{\emph{Comput. Phys. Commun.}
  {\bfseries 181} (2010) 1277}
  [\href{https://arxiv.org/abs/0906.0369}{{\ttfamily 0906.0369}}].

\bibitem{Arbey:2011zz}
A.~Arbey and F.~Mahmoudi, \emph{{SuperIso Relic v3.0: A program for calculating
  relic density and flavour physics observables: Extension to NMSSM}},
  \href{https://doi.org/10.1016/j.cpc.2011.03.019}{\emph{Comput. Phys. Commun.}
  {\bfseries 182} (2011) 1582}.

\bibitem{Arbey:2018msw}
A.~Arbey, F.~Mahmoudi and G.~Robbins, \emph{{SuperIso Relic v4: A program for
  calculating dark matter and flavour physics observables in Supersymmetry}},
  \href{https://doi.org/10.1016/j.cpc.2019.01.014}{\emph{Comput. Phys. Commun.}
  {\bfseries 239} (2019) 238}
  [\href{https://arxiv.org/abs/1806.11489}{{\ttfamily 1806.11489}}].

\bibitem{Altmannshofer:2008dz}
W.~Altmannshofer, P.~Ball, A.~Bharucha, A.J.~Buras, D.M.~Straub and M.~Wick,
  \emph{{Symmetries and Asymmetries of $B \to K^{*} \mu^{+} \mu^{-}$ Decays in
  the Standard Model and Beyond}},
  \href{https://doi.org/10.1088/1126-6708/2009/01/019}{\emph{JHEP} {\bfseries
  01} (2009) 019} [\href{https://arxiv.org/abs/0811.1214}{{\ttfamily
  0811.1214}}].

\bibitem{Descotes-Genon:2013vna}
S.~Descotes-Genon, T.~Hurth, J.~Matias and J.~Virto, \emph{{Optimizing the
  basis of $B\to K^*ll$ observables in the full kinematic range}},
  \href{https://doi.org/10.1007/JHEP05(2013)137}{\emph{JHEP} {\bfseries 05}
  (2013) 137} [\href{https://arxiv.org/abs/1303.5794}{{\ttfamily 1303.5794}}].

\bibitem{Hiller:2003js}
G.~Hiller and F.~Kruger, \emph{{More model-independent analysis of $b \to s$
  processes}}, \href{https://doi.org/10.1103/PhysRevD.69.074020}{\emph{Phys.
  Rev.} {\bfseries D69} (2004) 074020}
  [\href{https://arxiv.org/abs/hep-ph/0310219}{{\ttfamily hep-ph/0310219}}].

\bibitem{Paige:2003mg}
F.E.~Paige, S.D.~Protopopescu, H.~Baer and X.~Tata, \emph{{ISAJET 7.69: A Monte
  Carlo event generator for pp, anti-p p, and e+e- reactions}},
  \href{https://arxiv.org/abs/hep-ph/0312045}{{\ttfamily hep-ph/0312045}}.

\bibitem{Ellwanger:2005dv}
U.~Ellwanger and C.~Hugonie, \emph{{NMHDECAY 2.0: An Updated program for
  sparticle masses, Higgs masses, couplings and decay widths in the NMSSM}},
  \href{https://doi.org/10.1016/j.cpc.2006.04.004}{\emph{Comput. Phys. Commun.}
  {\bfseries 175} (2006) 290}
  [\href{https://arxiv.org/abs/hep-ph/0508022}{{\ttfamily hep-ph/0508022}}].

\bibitem{Allanach:2001kg}
B.C.~Allanach, \emph{{SOFTSUSY: a program for calculating supersymmetric
  spectra}}, \href{https://doi.org/10.1016/S0010-4655(01)00460-X}{\emph{Comput.
  Phys. Commun.} {\bfseries 143} (2002) 305}
  [\href{https://arxiv.org/abs/hep-ph/0104145}{{\ttfamily hep-ph/0104145}}].

\bibitem{Porod:2003um}
W.~Porod, \emph{{SPheno, a program for calculating supersymmetric spectra, SUSY
  particle decays and SUSY particle production at e+ e- colliders}},
  \href{https://doi.org/10.1016/S0010-4655(03)00222-4}{\emph{Comput. Phys.
  Commun.} {\bfseries 153} (2003) 275}
  [\href{https://arxiv.org/abs/hep-ph/0301101}{{\ttfamily hep-ph/0301101}}].

\bibitem{Djouadi:2002ze}
A.~Djouadi, J.-L.~Kneur and G.~Moultaka, \emph{{SuSpect: A Fortran code for the
  supersymmetric and Higgs particle spectrum in the MSSM}},
  \href{https://doi.org/10.1016/j.cpc.2006.11.009}{\emph{Comput. Phys. Commun.}
  {\bfseries 176} (2007) 426}
  [\href{https://arxiv.org/abs/hep-ph/0211331}{{\ttfamily hep-ph/0211331}}].

\bibitem{Eriksson:2009ws}
D.~Eriksson, J.~Rathsman and O.~Stal, \emph{{2HDMC: Two-Higgs-Doublet Model
  Calculator Physics and Manual}},
  \href{https://doi.org/10.1016/j.cpc.2009.09.011}{\emph{Comput. Phys. Commun.}
  {\bfseries 181} (2010) 189}
  [\href{https://arxiv.org/abs/0902.0851}{{\ttfamily 0902.0851}}].

\bibitem{Arbey:2016kqi}
A.~Arbey, S.~Fichet, F.~Mahmoudi and G.~Moreau, \emph{{The correlation matrix
  of Higgs rates at the LHC}},
  \href{https://doi.org/10.1007/JHEP11(2016)097}{\emph{JHEP} {\bfseries 11}
  (2016) 097} [\href{https://arxiv.org/abs/1606.00455}{{\ttfamily
  1606.00455}}].

\bibitem{Bhom:2020lmk}
J.~Bhom, M.~Chrzaszcz, F.~Mahmoudi, M.T.~Prim, P.~Scott and M.~White, \emph{{A
  model-independent analysis of $b \to s \mu^{+} \mu^{-}$ transitions with
  GAMBIT's FlavBit}},  \href{https://arxiv.org/abs/2006.03489}{{\ttfamily
  2006.03489}}.

\bibitem{Workgroup:2017myk}
{\scshape GAMBIT Flavour Workgroup} collaboration, \emph{{FlavBit: A GAMBIT
  module for computing flavour observables and likelihoods}},
  \href{https://doi.org/10.1140/epjc/s10052-017-5157-2}{\emph{Eur. Phys. J.}
  {\bfseries C77} (2017) 786}
  [\href{https://arxiv.org/abs/1705.07933}{{\ttfamily 1705.07933}}].

\bibitem{Chobanova:2017ghn}
V.G.~Chobanova, T.~Hurth, F.~Mahmoudi, D.~Martinez~Santos and S.~Neshatpour,
  \emph{{Large hadronic power corrections or new physics in the rare decay $B
  \to K^{*}\mu^{+}\mu^{-}$?}},
  \href{https://doi.org/10.1007/JHEP07(2017)025}{\emph{JHEP} {\bfseries 07}
  (2017) 025} [\href{https://arxiv.org/abs/1702.02234}{{\ttfamily
  1702.02234}}].

\bibitem{Arbey:2018ics}
A.~Arbey, T.~Hurth, F.~Mahmoudi and S.~Neshatpour, \emph{{Hadronic and New
  Physics Contributions to $b \to s$ Transitions}},
  \href{https://doi.org/10.1103/PhysRevD.98.095027}{\emph{Phys. Rev.}
  {\bfseries D98} (2018) 095027}
  [\href{https://arxiv.org/abs/1806.02791}{{\ttfamily 1806.02791}}].

\bibitem{Hurth:2020rzx}
T.~Hurth, F.~Mahmoudi and S.~Neshatpour, \emph{{Implications of the new LHCb
  angular analysis of $B \to K^* \mu^+ \mu^-$ : Hadronic effects or new
  physics?}}, \href{https://doi.org/10.1103/PhysRevD.102.055001}{\emph{Phys.
  Rev.} {\bfseries D102} (2020) 055001}
  [\href{https://arxiv.org/abs/2006.04213}{{\ttfamily 2006.04213}}].

\bibitem{Arbey:2019duh}
A.~Arbey, T.~Hurth, F.~Mahmoudi, D.M.~Santos and S.~Neshatpour, \emph{{Update
  on the $b \to s$ anomalies}},
  \href{https://doi.org/10.1103/PhysRevD.100.015045}{\emph{Phys. Rev.}
  {\bfseries D100} (2019) 015045}
  [\href{https://arxiv.org/abs/1904.08399}{{\ttfamily 1904.08399}}].

\bibitem{Hurth:2020ehu}
T.~Hurth, F.~Mahmoudi and S.~Neshatpour, \emph{{Model independent analysis of
  the angular observables in $B^{0} \to K^{*0} \mu^+ \mu^-$ and $B^{+} \to
  K^{*+} \mu^+ \mu^-$}},  \href{https://arxiv.org/abs/2012.12207}{{\ttfamily
  2012.12207}}.

\bibitem{Aaij:2020nrf}
{\scshape LHCb} collaboration, \emph{{Measurement of $CP$-Averaged Observables
  in the $B^{0}\rightarrow K^{*0}\mu^{+}\mu^{-}$ Decay}},
  \href{https://doi.org/10.1103/PhysRevLett.125.011802}{\emph{Phys. Rev. Lett.}
  {\bfseries 125} (2020) 011802}
  [\href{https://arxiv.org/abs/2003.04831}{{\ttfamily 2003.04831}}].

\bibitem{Aaij:2020ruw}
{\scshape LHCb} collaboration, \emph{{Angular analysis of the $B^{+}\rightarrow
  K^{\ast+}\mu^{+}\mu^{-}$ decay}},
  \href{https://arxiv.org/abs/2012.13241}{{\ttfamily 2012.13241}}.

\bibitem{Mahmoudi:2014mja}
F.~Mahmoudi, S.~Neshatpour and J.~Virto, \emph{{$B \to K^{*} \mu^{+} \mu^{-}$
  optimised observables in the MSSM}},
  \href{https://doi.org/10.1140/epjc/s10052-014-2927-y}{\emph{Eur. Phys. J.}
  {\bfseries C74} (2014) 2927}
  [\href{https://arxiv.org/abs/1401.2145}{{\ttfamily 1401.2145}}].

\bibitem{Uhlrich:2020ltd}
G.~Uhlrich, F.~Mahmoudi and A.~Arbey, \emph{{MARTY -- Modern ARtificial
  Theoretical phYsicist: A C++ framework automating symbolic calculations
  Beyond the Standard Model}},
  \href{https://arxiv.org/abs/2011.02478}{{\ttfamily 2011.02478}}.

\end{thebibliography}\endgroup

\end{document}